# Development and validation of Critical Thinking Disposition Inventory for Chinese medical college students (CTDI-M)

*Xiaoxia Wang[1,*], Xiaoxiao Sun[2], Tianhao Huang[1], Renqiang He[1], Weina Hao[1], Li Zhang[1]*

[1] Department of Basic Psychology, School of Psychology, Army Medical University, Chongqing 400038, China

2 Department of Psychological Nursing, School of Nursing, Army Medical University, Chongqing 400038, China

**Abstract PURPOSE**: Critical thinkers in medical context must be not only "able" but also "willing" to think critically. To develop and conduct psychometric testing of Critical Thinking Disposition Inventory which measure the critical thinking disposition of Chinese medical college students. **METHODS**: The study was conducted in two stages: (a) item generation and exploratory factor analysis (EFA) and (b) testing of psychometric properties (*construct validity*, *internal consistency reliability* (split-half reliability and Cronbach' s alpha), and *test-retest reliability*). The subjects were composed of 441 undergraduate medical and nursing students from a medical university in China. Test-retest reliability of the instrument was determined at two-week interval. Data was analyzed with SPSS13.0. **RESULTS**: Preliminary 264 items were obtained using an open-ended questionnaire; from which 61 items were reviewed through half open-ended questionnaire, and finally 18 items were chosen. Eighteen final items were sorted into 3 factors, which were identified as "*Open-mindedness* (7 items)", "*Systematicity/analyticity* (6 items)" and "*Truth seeking* (5 items)". The cumulative percent of variance was 57.66%. The reliability of the scale (Cronbach's alpha) was 0.924 and the factors' alphas ranged from 0.824-0.862. Correlational analysis indicated moderate to high correlations between the subscales and total scores of the CTDI-CM. Our results indicated that open-mindedness and systematicity/analyticity were higher for medical students than nursing students. **CONCLUSIONS**: This study conducted in a Chinese medical college student population demonstrated a reliable and valid instrument for clinical thinking disposition, which measured motivation and cognitive components. The effect of enrollment year and major on the profiles of critical thinking dispositions was identified, emphasizing the importance of applying specialized teaching to students of different majors.

**Keywords**：Critical thinking disposition; California Critical Thinking Disposition Inventory (CCTDI); medical college students; validity; reliability

## Introduction

Critical thinking is needed more than ever to become an adaptive and flexible learner in the information age (Dwyer, Hogan, & Stewart, 2014). The importance of being 'critical' among medical students and practitioners have also been increasingly emphasized (Gupta & Upshur, 2012). The Delphi Report announced in 1990 by the theoreticians of the USA and Canada defined critical thinking (CT) as the ability to apply cognitive skills (interpretation, analysis, inference, evaluation, explanation, and self-regulation) and the disposition toward CT (being open-minded or intellectually honest) (Athari, Sharif, Nasr, & Nematbakhsh, 2013; Papp et al., 2014). Critical thinking was frequently referred to as the individuals' cognitive ability to think and make correct decisions independently, and utilizing rational/logical thought (Harasym, Tsai, & Hemmati, 2008; Shirkhani, 2011).

Increasing attention has been paid to the individual differences in critical thinking disposition, which is defined as the tendency or attitude to understand the need for a particular skill and the willingness to make the effort in applying it (Aizikovitsh-Udi & Radakovic, 2012), or to simply put it, the attitude toward critical thinking. Dispositions toward critical thinking are vital to critical-thinking performance (Hajhosseiny, 2012) and professional clinical judgment (Jenicek, Croskerry, & Hitchcock, 2011). Practically, both disposition and ability are necessary for critical thinking and that neither is likely to be sufficient alone (Krupat et al., 2011). The assessment of CT dispositions may help identify the target to promote critical thinking through training programs both in professional and education context.

---

[1] **Corresponding author**: Li Zhang, **Tel**: +86-23-68771770; **E-mail**: zhangli-372@163.com; **Address**: Department of Basic Psychology, School of Psychology, Amy Medical University, Chongqing 400038, China.

The most widely used measurement tool in China was the translated version of California Critical Thinking Disposition Inventory (CCTDI) (Peng, Wang, & Chen, 2004; Yeh, 2002). The CCTDI is calibrated for use with the general adult population including workers and working professionals at all levels and students in grades 10 and above. CCTDI included seven subscales: "*Inquisitiveness*", "*Systematicity*", "*Analyticity*", "*Truth seeking*", "*Open-mindedness*", "*Self-confidence*", and "*Maturity*" (Facione, Facione, & Sanchez, 1994). Yeh et al. translated the CCTDI into Chinese, and administered it to a nursing undergraduate student sample in Taiwan. Compared to the English CCTDI (alpha=0.79), the overall alpha (0.71) of Chinese CCTDI was inferior, and internal consistencies (Cronbach α) of three subscales were inadequate (*open-mindedness=0.34, analyticity=0.40, and systematicity=0.47*) (Yeh, 2002). In addition, the content validity of these three subscales were moderate (CVI=.50 to .67), as compared to the English CCTDI subscales (CVI=.82 to 1). Peng et al. developed a conceptually equivalent version of CCTDI: CTDI-CV (Critical Thinking Disposition Inventory-Chinese Version), which showed more satisfactory subscale alphas ranging from 0.54 to 0.77, and overall alpha of 0.90. CVIs for "*open-mindedness*", "*analyticity*" and "*systematicity*" subscales were improved to .90-1. However, Cronbach α (*Chinese CCDTI=.46, CTDI-CV=.57*) and CVI (*Chinese CCDTI=.70, CTDI-CV=.60*) of "*maturity*" subscale for both Chinese CCTDI and CTDI-CV were lower than English CCTDI (alpha=.64, CVI=.90) (Peng et al., 2004). For another translated Chinese version of CCTDI for university students, Cronbach α of "*open-mindedness*" (0.39), *"systematicity"* (0.43) and "*maturity*" (0.45) subscales were also not satisfactory (Luo & Yang, 2001). Collectively, the conceptualizations and measurement of CT dispositions in the Chinese-speaking population merit further exploration.

An important factor which may explain the diversity of psychometric characteristics of versions of CCTDI is culture sensitivity. According to a literature review, Asian students tend to show less critical thinking dispositions compared with non-Asian countries (Jenkins, 2011; Li, Li, & Lù, 2006; Salsali, Tajvidi, & Ghiyasvandian, 2013). Specifically, (1) Analyticity and systematicity are the cognitive components of CT dispositions which are culture-sensitive. "Analyticity" means the use of evidence and anticipation of possible consequences to resolve problems. "Systematicity" means being organized, focused and diligent in resolving problems. In comparison with American university students, the percentage of students with lower than moderate level of "Analyticity" and "Systematicity" is greater in Chinese population (Luo & Yang, 2001). Specifically for Chinese medical undergraduates, the average score of "Systematicity" were at the lowest level of all subscales of CT dispositions (Xie, Cao, Zhang, Zhu, & Zhang, 2014). The systematicity of Chinese nursing students was of moderate level (Peng et al., 2004). Previous study suggested that western cultures tend to be analytic, whereas the traditional Chinese societies tend to be holistic and synthetic which was manifested in language (Tang, 2004), thinking models of medicine (Wang et al., 2011) and preferences for dialectical proverbs and dialectical resolution of social contradiction (Nisbett, 2005). Since the cognitive components of CT dispositions is cruicial to effective critical thinking, cultural differences in thinking patterns need to be considered in the context of Chinese culture..

(2) "Inquisitiveness", "Truth seeking" and "Open-mindedness" are the motivation components of CT dispositions. CT ability is presumed to be significantly correlated with motivation towards learning (Dwyer, Hogan, & Stewart, 2014). First, inquisitiveness refers to the inclination to be curious and eager to learn the knowledge that even may not be of immediate use. This may encourage learners to engage in deep and creative reasoning (Albergaria-Almeida, 2011). Second, the average score of truth-seeking were at the lowest level of all subscales of CT dispositions among Chinese medical undergraduates (Xie et al., 2014). The truth-seeking of Chinese nursing students were of moderate level (Peng et al., 2004). Asian university students tended to learn for pragmatic purposes compared to American university students who tend to possess the attitude towards valuing truth (Peng & Nisbett, 1999). While open-mindedness and truth-seeking have been deemed as important in good critical thinkers, only truth-seeking significantly predicts Chinese students' critical thinking performance, and those answers concerned more about seeking for solutions from authorities or preconception, rather than seeking independent evidence or reasoning (Ku & Ho, 2010). Third, in comparison with American university students, the proportion of students displaying lower than moderate level of open-mindedness is greater in Chinese population (Luo & Yang, 2001). Open-minded people in Asian culture may be more inclined to accept contradictory propositions and avoiding social conflicts. These culture diversities may explain the low internal consistencies of

"Open-mindedness" for CCTDI in Chinese nursing students (Yeh, 2002). Therefore, the motivation components of CT dispositions need to focus on these culture-sensitive traits and examine how these traits may influence medical performance.

(3) "Self-confidence" and "Maturity" are the personality components of CT dispositions. The self-confidence scale measures the trust one places in one's own reasoning processes. Actually, emotionally taxing situations, threats to individual identity (e.g. gender prejudice) or inappropriate priorities of values will challenge self-confidence of critical thinkers and impair the ability to self-reflect (Papp et al., 2014), thus self-confidence is particularly important to CT dispositions. The maturity scale targets the disposition to be judicious in one's decision-making, and thus require self-reflection which develops gradually from adolescence to adulthood. The students who exhibit lower than moderate level of maturity occupy a bigger proportion compared with American university students (Luo & Yang, 2001). Similarly, more research is needed to develop the instruments specialized for Chinese context.

Empirically. critical thinking is valued both for nursing (Spencer, 2008) and clinical expertise (Harasym et al., 2008). Critical thinking should improve diagnostic skills and reduce errors in management (Harasym et al., 2008). Critical thinking not only constitutes the ability to think in accordance with the rules of logic and probability, but also the capability to solve problems which are content-dependent (Aizikovitsh & Amit, 2010). For instance, transfer of knowledge and skills obtained from the classroom to the clinical environment requires for critical thinking skill and clinical judgment (Kermansaravi, Navidian, & Kaykhaei, 2013). As a result, CT skills have been emphasized by Global Minimum Essential Requirements as one of the seven student competence domains which included the following: (1) Professional Values, Attitudes, Behavior and Ethics; (2) Scientific Foundation of Medicine; (3) Clinical Skills; (4) Communication Skills; (5) Population Health and Health Systems; (6) Management of Information; (7) *Critical Thinking and Research* (Shiau & Chen, 2008). The competences contained in the "GMER" define what a physician is (Wojtczak & Schwarz, 2003). These domains have been assessed with objective structured clinical examination (OSCE), in which the competence of critical thinking and research could be defined as the ability to formulate hypotheses, collect and critically evaluate data for the solution of problems (Stern, Wojtczak, & Schwarz, 2002). Although critical thinking competencies are generic abilities, CT behaviors cannot be effectively learned or taught without a discipline-related practice setting (Kenimer, 2002). Therefore, knowledge about the individual differences of CT dispositions specific to medical discipline could facilitate teaching critical thinking. In contrast with the two Chinese versions of CCTDI which were directed at nursing students only, the current study also included medical students during development of CT disposition assessment tool.

Collectively, an instrument for measuring critical thinking dispositions developed independently for Chinese medical students promises to: (1) increase the content validity of specific factors of CT disposition (i.e. open-mindedness, analyticity, systematicity and maturity); and (2) identify those traits with greater culture differences and evaluate the criterion validity of their measure.

**2 Method**

***2.1 Study 1: Development and validation of Critical Thinking Disposition Inventory for Chinese medical college students (CTDI-M)***

*2.1.1 Participants*

During the *first* stage (*item generation and content validity*), 161 clinical medicine undergraduate students, 10 educational specialists and 10 psychological specialists were recruited and completed an open-ended questionnaire. From the original 181 surveys distributed, 177 surveys were deemed valid and analyzed (average age: 23.50±9.76; age range: 18-62). Among this sample 157 subjects being males (average age: 22.03±7.84) and 20 being female (average age: 23.45±8.41). The response rate was 97.79%.

To complete a semi-opened questionnaire, 138 clinical medicine undergraduate students, 61 nursing undergraduate students were recruited, together with 10 educational specialists and 10 psychological specialists. From the original 219 surveys distributed, 209 surveys were deemed valid and analyzed (average age: 23.36±9.03; age range: 18-62). Among this sample 99 subjects being males (average age: 23.26±9.72) and 110 being female (average age: 23.45±8.41). The response rate was 95.43%.

After that, 299 undergraduate students in clinical medicine, 71 students in preventive medicine or medical laboratory science and 61 students in nursing, together with 10 educational specialists and 10 psychological specialists, were recruited and completed a closed-ended questionnaire of CT dispositions. From the original 451 surveys distributed, 442 surveys were deemed valid and analyzed (average age: 21.97±6.37; age range: 17-62). Among this sample 342 subjects being males (average age: 21.35±5.41) and 100 being female (average age: 24.08±8.61). The response rate was 98.00%.

During the *second* stage (*testing of psychometric properties*), a cross-sectional survey research was conducted with an anonymous questionnaire of critical thinking disposition. Participants were 441 undergraduate students (323 males and 97 females) at a medical university in China (95.24% response rate). The resulting convenience sample consisted of 61 students major in nursing, 278 students in clinical medicine, 41 students in preventive medicine and 40 students in medical laboratory science. The average age of the participants was 20.64 years (S.D.=1.19; range=17-25). This sample size had enough power to detect significant difference revealed by power analysis.

Verbal informed consent was obtained from each participant prior to surveys. All information gained from the study participants was confidential and participants could withdraw from the study at any time on a voluntary basis.

### *2.1.2 Statistics*

We first designed an open-ended questionnaire ("What are the aspects of critical thinking disposition for medical college students?") to investigate the components of critical thinking disposition. Preliminary 264 items were obtained. Based on the conceptual framework from literature review, and the results of open-ended questionnaire, 97 items were extracted and entered into a half-opened questionnaire (Appendix 1). Those items (a total of 61) endorsed by more than 50% of respondents were obtained and entered into a closed-ended questionnaire (Appendix 2). Participants were required to complete the questionnaire on a five-point Likert-type scale (1=disagree strongly; 2=disagree somewhat; 3=neutral; 4=agree somewhat; 5=agree strongly). Then we performed exploratory factor analysis (EFA) to explore the structure of the instrument, using principle component analysis with an orthogonal (varimax) rotation solution and a minimum eigenvalue of 1 as the cut-off point for the totaled factors.

For internal consistency, split-half reliability was computed, and Cronbach's alpha coefficient was calculated for each domain and global score. Test-retest reliability was analyzed by the intra-class correlation coefficients (ICCs), with a two-week interval between administrations to the same group, which was long enough to avoid confounding effects of practice and allow for a natural change in the construct. Construct validity was also analyzed. Pearson's correlation was used in the construct validity, comparing each subscale's score to the overall score. The criterion validity was obtained by computing the inter-correlations among the scores of subscales of CTDI-CV and CTDI-CM.

### *2.2 Study 2: Comparisons of Critical Thinking Disposition among groups of medical students (CTDI-M)*

### *2.2.1 Participants*

Three subgroups of participants were recruited and completed the CTDI-CM shortly after the new term began. These included 641 medical graduate students (average age: 26.73±3.96; age range: 21-41), 420 medical graduate students (enrolled in 2012) (average age: 20.64±1.19; age range: 17-25) and 289 medical undergraduate students (enrolled in 2014) (average age: 19.57±1.63; age range: 17-25). The gender ratio is unbalanced between the subgroups (graduate students: male: 397, female: 244; undergraduate students 2012: male: 266, female: 23; undergraduate students 2012: male: 333, female: 87), and thus gender was entered into the linear regression model as a covariate.

*2.2.2 Statistics*

The multivariate linear regression model was utilized to estimate the significant predictive factor of CT dispositions, with major and subgroup (enrollment year of college) as independent variables and age as covariate.

**3 Results**

*3.1 Questionnaire and the factor analysis*

We obtained the 18-item questionnaire regarding Chinese medical college students' critical thinking dispositions (Table 1). The subjects were asked "To what extent do you agree that this phrase correctly describes you?" when they completed the questionnaire.

**Table 1 The initial measurement indicators of medical students' critical thinking dispositions**

*Instructions: Please take your time and read each question carefully before answering it. For multiple choice questions, place a √ in the box next to the ONE BEST answer.*

| initial measurement indicators | 1 disagree strongly | 2 disagree somewhat | 3 neutral | 4 agree somewhat | 5 agree strongly |
|---|---|---|---|---|---|
| **1. Fair and objective attitude**(公正客观对待事物) | | | | | |
| **2. Avoiding existing cognition to hinder my judgments**(不使原有认识阻碍判断) | | | | | |
| **3. Seeking evidence**(寻求证据) | | | | | |
| **4. Accepting different views**(接纳不同观点) | | | | | |
| **5. Seeking solutions from many aspects**(寻求多样性答案) | | | | | |
| **6. Finding the truth**(寻找真相) | | | | | |
| **7. Making a decision wisely and prudently**(明智和谨慎地做决定) | | | | | |
| **8. Avoiding the negative effect of mental set**(不陷于惯性思维定式) | | | | | |
| **9. In-depth thinking**(能够抓住事情深层次问题) | | | | | |
| **10. Making a comprehensive analysis of a problem actively** (能动、全面分析事物的各方面) | | | | | |
| **11. Breaking habitual thinking patterns**(打破思维习惯) | | | | | |
| **12. Logic thinking**(逻辑思维) | | | | | |
| **13. Active thinking**(主动思考) | | | | | |
| **14. Avoiding occasional indication to waver my thinking**(不受偶然的暗示而犹豫动摇) | | | | | |
| **15. Distinguishing truth from falsehood**(在表面理由背后挖掘真相) | | | | | |
| **16. No blind faith in authority**(不迷信权威) | | | | | |
| **17. Viewing the problem in many ways**(多方面审视问题) | | | | | |
| **18. Sifting the true from the false** (去伪存真) | | | | | |

The factor analysis identified a three-factor model, with the eigenvalues of 3 factors exceeded one, accounting for 57.66% of extraction sums of squared loadings (Table 2). Kaiser-Meyer-Olkin measure of sampling adequacy was 0.943, which was greater than 0.5 for a satisfactory factor analysis to proceed. Bartlett's Test of Sphericity ($x^2$=3725.54, $P$<0.000) suggested the appropriateness of the factor analysis model.

The first factor (open-mindedness) meant the open attitude and willingness to listen to and consider other people's ideas and suggestions before arriving at conclusions. The second factor (systematicity/ analyticity) meant the trait of being painstaking, careful and effective decision-making and problem-solving. The third factor (truth seeking) meant the state of active curiosity, active engagement in thinking, and avoiding the negative effect of mental set.

**Table 2 Factor Loadings of Each Item of CTDI-CM (Critical Thinking Disposition Inventory for Chinese medical students)**

| Measurement indicators (Factors) | Items | Rotated sums of squared loadings | | Rotated component matrix(α) | | |
|---|---|---|---|---|---|---|
| | | % of variance | Cumulative % | Factor 1 | Factor 2 | Factor 3 |
| **Open-mindedness** | 5. Accepting different views(接纳不同观点) | 20.07% | 20.07% | 0.792 | | |
| | 2. Fair and objective attitude(公正客观对待事物) | | | 0.721 | | |
| | 8. Making a decision wisely and prudently(明智和谨慎地做决定) | | | 0.689 | | |
| | 11. Making a comprehensive analysis of a problem actively (能动、全面分析事物的各方面) | | | 0.606 | | |
| | 14. Active thinking(主动思考) | | | 0.605 | | |
| | 19. Eliminating the false and retaining the true(去伪存真) | | | 0.591 | | |
| | 17. No more blind faith in authority(不迷信权威) | | | 0.549 | | |
| **Systematicity/ analyticity** | 7. Finding the truth(寻找真相) | 19.57% | 39.64% | | 0.768 | |
| | 4. Seeking evidence(寻求证据) | | | | 0.735 | |
| | 13. Logic thinking(逻辑思维) | | | | 0.646 | |
| | 16. Distinguish truth from falsehood(在表面理由背后挖掘真相) | | | | 0.633 | |
| | 18. Viewing the problem in many ways(多方面审视问题) | | | | 0.610 | |
| | 10. In-depth thinking(能够抓住事情深层次问题) | | | | 0.579 | |
| **Truth seeking** | 9. Avoiding the negative effect of mental set(不陷于惯性思维定式) | 18.02% | 57.66% | | | 0.786 |
| | 12. Breaking habitual thinking patterns(打破思维习惯) | | | | | 0.732 |
| | 3. Existing cognition not to obstruct my judgments(不使原有认识阻碍判断) | | | | | 0.707 |
| | 15. Unrealistic suggestions not to waver my thinking(不受偶然的暗示而犹豫动摇) | | | | | 0.667 |
| | 6. Seeking solutions from many aspects(寻求多样性答案) | | | | | 0.522 |

Note. extraction method: principal component analysis; rotated method: varimax with Kaiser normalization

### 3.2 Reliability analysis

Internal consistency reliability (Cronbach's alpha coefficient), split-half reliability and test-retest reliability were analyzed (Table 3). Results showed that the correlations between each subscale score and the total CDTI-CM score were all statistically significant, with Conbach's alpha ranging from 0.776 to 0.965. The questionnaire had substantial reliability according to the following criterion: 0.9 as excellent, 0.8 as good, 0.7 as acceptable, 0.6 as questionable, 0.5 as poor and less than 0.5 as unacceptable. Furthermore, the split-half reliability of CDTI-CM ranged from 0.776 to 0.922, and the two-week test-retest reliability of CDTI-CM was 0.808 to 0.965.

Table 3 Reliability analysis of CTDI-CM (Critical Thinking Disposition Inventory for Chinese medical students)

| Reliability | coefficient | | | |
| --- | --- | --- | --- | --- |
| | Measurement indicators | Open-mindedness (Factor 1) | Systematicity/ analyticity (Factor 2) | Truth seeking (Factor 3) |
| Internal consistency reliability | 0.924* | 0.862* | 0.848* | 0.824* |
| Split-half reliability | 0.922* | 0.812* | 0.827* | 0.776* |
| Two-week test-retest reliability | 0.881* | 0.808* | 0.965* | 0.907* |

*$p<0.01$ (2-tailed), Pearson correlation

### 3.3 Validity analysis

The construct validity was quantified by verifying a construct shared by the subscales. The correlation coefficients between subscale and overall scores varied from 0.851 to 0.901. The correlation coefficients between each pair of subscales were 0.630-0.692. Thus the correlations between subscales with the overall score were higher than those found among the subscales, which indicates that the scales share an adjacent construct (Table 4).

Table 4 Correlation coefficients between variables of CTDI-CM (Critical Thinking Disposition Inventory for Chinese medical students)

| Variables (number of items) | 95% confidence interval | | | |
| --- | --- | --- | --- | --- |
| | Measurement indicators | Open-mindedness (Factor 1) | Systematicity/ analyticity (Factor 2) | Truth seeking (Factor 3) |
| Measurement indicators(18) | 1 | 0.901* | 0.882* | 0.851* |
| Open-mindedness(7) | | 1 | 0.692* | 0.630* |
| Systematicity/ analyticity (6) | | | 1 | 0.655* |
| Truth seeking (5) | | | | 1 |

*$p<0.01$ (2-tailed), Pearson correlation

The criterion validity was determined by correlating subscales of CTDI-CM with CTDI-CV. Open-mindedness of CTDI-CM was positively correlated with maturity and self-confidence of CTDI-CV (r=0.170, $P<0.01$). Systematicity/analyticity of CTDI-CM was positively correlated with self-confidence (r=0.215, $P<0.01$) of CTDI-CV, and negatively correlated with inquisitiveness of CTDI-CV (r=-0.219, $P<0.01$). Truth seeking of CTDI-CM was positively correlated with self-confidence of CTDI-CV (r=0.200, $P<0.01$), and negatively correlated with inquisitiveness (r=-0.318, $P<0.01$) and analyticity (r=-0.129, $P<0.05$) of CTDI-CV. However, there was no significant correlation between the total scores of CTDI-CV and CTDI-CM (r=0.028, $P=0.643$) (Table 5).

Table 5 Inter-correlations among the scores of subscales of CTDI-CV and CTDI-CM[a]

| | | CTDI-CV | | | | | | | CTDI-CM | | |
|---|---|---|---|---|---|---|---|---|---|---|---|
| | | Open-Mindedness[a] | Systematicity[a] | Inquisitiveness | Maturity | Analyticity | Truth-seeking[a] | Self-confidence | Open-Mindedness[b] | Systematicity/analyticity | Truth seeking[b] |
| CTDI-CV | Open-mindedness[a] | 1 | .129* | -.062 | .066 | -.287** | .156** | -.116 | -.044 | -.033 | .058 |
| | Systematicity[a] | | 1 | .091 | .334** | -.110 | .315** | -.411** | .039 | -.102 | -.061 |
| | Inquisitiveness | | | 1 | -.136* | .279** | -.161** | -.103 | -.115 | -.219** | -.318** |
| | Maturity | | | | 1 | -.306** | .635** | -.563** | .170** | .062 | .075 |
| | Analyticity | | | | | 1 | -.232** | .077 | -.092 | -.104 | -.129* |
| | Truth-seeking[a] | | | | | | 1 | -.563** | .031 | -.031 | -.003 |
| | Self-confidence | | | | | | | 1 | .140* | .215** | .200** |
| CTDI-CM | Open-mindedness[b] | | | | | | | | 1 | .638** | .606** |
| | Systematicity/analyticity | | | | | | | | | 1 | .689** |
| | Truth seeking[b] | | | | | | | | | | 1 |

Notes: *p < .05; **p < .01. a: CTDI-CV (Critical Thinking Disposition Inventory-Chinese Version). b: CTDI-CM (Critical Thinking Disposition Inventory for Chinese medical students)

*3.4 Predictive factors of critical thinking dispositions: major, age and enrolled year*

(1) Comparison among majors: Independent sample t-test between medical and nursing students revealed that open-mindedness and systematicity/analyticity mean scores were higher for medical students than nursing students (Table 6).

Table 6 Comparison of mean Scores of Critical Thinking Dispositions between medical and nursing students (mean±SD)

| Factor(number of items) | discipline | |
| --- | --- | --- |
|  | medical students | nursing students |
| **Open-mindedness** | 4.07(0.58)[a] | 3.82(0.67) |
| **Systematicity/analyticity** | 4.05(0.55)[a] | 3.82(0.58) |
| **Truth seeking** | 3.84(0.66) | 3.73(0.57) |

\* p<0.01, (2-tailed), independent sample t-test, compared with nursing students

(2) Comparison among age groups: The general linear model (GLM) indicated that age did not significantly predict CT dispositions of medical students, while major and subgroup significantly predicted CT dispositions of medical students ($P$=0.404, partial $\eta^2$=0.004). However, *Pearson* correlation between age and "Open-mindedness" revealed significant negative associations (r=0.06, $P$=0.02).

(3) Comparison among subgroups with different enrolled year: The subgroup (A: undergraduate 2012, B: graduate 2014, C: undergraduate 2014) indicated different enrolled year of the students. Post-hoc comparisons between each two subgroups of CT dispositions revealed that "Open-mindedness" scores of undergraduate 2012 were significantly greater than graduate 2014 and undergraduate 2014 ($P$s<0.001). *Systematicity/analyticity* and *truth seeking* scores of undergraduates 2012 were significantly greater than those of graduates 2014 ($P$<0.001). *Truth seeking* scores of graduates 2014 were significantly less than undergraduates 2014 ($P$s<0.001).

**4 Discussions**

The study emphasized the importance of developing a reliable and valid CT disposition instrument appropriate for use in Chinese cultures. To our knowledge, this is the first study to develop a questionnaire designed for evaluating the critical thinking dispositions of medical students in China. The primary goal of the present study was to identify the factor structure of the CTDI-CM, and test the reliability and validity of the instrument. Critical thinking dispositions of Chinese medical college students include open-mindedness, systematicity/ analyticity and truth-seeking. Open-mindedness represents an open attitude underlies the willingness to consider different viewpoints and options before arriving at conclusions. Systematicity/ analyticity holds for values such as fairness and truth and the skills striving for sound and unbiased judgments. Truth-seeking refers to an interest in or enjoyment of thinking and is a prerequisite for active engagement in thinking. The present study revealed that Chinese version of CDTI-CM for medical college students showed acceptable psychometric properties.

**4.1 Comparison among different versions of inventory for critical thinking dispositions**

Since non-cognitive factors may have great impact on the participants' critical thinking, such as culture and motivation (Fábián, 2015), we compared the psychometric properties of CTDI-CM with Chinese CCTDI (California Critical Thinking Dispositions Inventory) (Yeh, 2002) and CTDI-CV (Critical Thinking Disposition Inventory-Chinese Version) (Peng et al., 2004), which were developed conceptually or semantically to be equivalent to the original CCTDI respectively. While the Chinese CCTDI verified construct validity for *truth-seeking, open-mindedness, systematicity* and *maturity* subscale (Yeh, 2002), the content validity (alpha = 0.34) of *open-mindedness* for Chinese CCTDI was less satisfactory than the current study (alpha = 0.86). Additionally, criterion validity analysis revealed that open-mindedness of CTDI-CV was irrelevant to the three factors of CTDI-CM. Thus the content validity of CTDI-CM is different from CTDI-CV. *Open-mindedness* (CTDI-CM) means to be open to divergent views, to be prudent in decision making, and most importantly, not to be submissive to authorities. By contrast, *open-mindedness* (CTDI-CV) addresses being tolerant of divergent world views/cultures and readiness to monitor one's own cognitive bias. The attitude towards authorities may differentiate the content of the two measurements. An indirect evidence was that Chinese undergraduate nursing students are not as open-minded as their counterparts of American students (Yeh & Chen, 2003), and have ambivalent attitudes toward this disposition (Tiwari, Avery, & Lai, 2003). Chinese students as obedient learners may be more submissive to their teachers and dependent on rules, which may hinder their willingness to be open-minded.(Li & Wegerif, 2014). Consequently, the inclination of open-minded may lead to more solid decision-related reasoning and prohibit the nurses to from medical errors when they implement the clinical decision by doctors. Therefore, the content of open-mindedness factor of CTDI-CM may be more suitable to detect the potential inclination of those Furthermore, open-mindedness of CTDI-CM is significantly correlated with *maturity* of CTDI-CV. A plausible explanation was that open-mindedness may depend on self-reflection which show different developmental trajectories between young adults of different cultures. However, the developmental characteristics of open-mindedness merit further explorations.

In addition, *systematicity/analyticity* and *truth seeking* of CTDI-CM were positively related to self-confidence of CTDI-CV, while negatively *related to inquisitiveness* of CTDI-CV. The latter result may reflect the different emphases of *inquisitiveness* (CTDI-CV) and *truth-seeking* (CTDI-CM). The *inquisitiveness* stands for eagerness to explore the unknowns and interest in mechanisms behind the phenomenon, while *truth-seeking* and *systematicity/analyticity* measures the cognitive operations following informal and formal logical rules. Therefore, the motivation aspect of CT dispositions was less emphasized in CTDI-CM than CTDI-CV, which explained the inverse relationship between the two groups of factors. Meanwhile, Chinese philosophy and Confucius' teaching emphasize thinking as reflection in the context of relationships and identification with the interests of the whole (Li & Wegerif, 2014), which may help explain the negative relationship between *inquisitiveness* (CTDI-CV) and *systematicity/analyticity* (CTDI-CM).

Compared with previous study (Yeh, 2002), test-retest analysis of CTDI for medical students yielded more stable results across two assessment occasions (2 weeks apart) in current study, with all correlations statistically significant, ranging from 0.808 to 0.965 with an overall correlation of 0.881. And the results also supported the internal consistency reliability of the Chinese version of CTDI for medical students, which performed better than the two Chinese versions of CCTDI (Peng et al., 2004; Yeh, 2002). These results confirm the current inventory as a reliable instrument for measuring critical thinking disposition.

**4.2 The predictors of CT dispositions: age, major and enrollment year of college**

An interesting and counter-intuitive finding was that increasing age may compromise the open-mindedness CT disposition. The results were replicated when considering enrollment year of college as predictor of CT disposition.

In health care disciplines, medical and nursing may both require high critical thinking dispositions which could lead to increased quality of care and better treatment outcomes. Our results suggested that medical students performed better on open-mindedness and systematicity/ analyticity than nursing students, which contrast with previous findings. A survey conducted with Chinese version of CTDI (CTDI-CV) indicated that general performance of critical thinking ability in medical and nursing was positive (overall score>280) (Peng et al., 2004). Dispositional differences using CCTDI among several majors [(practice disciplines: i.e., nursing, education, business) and nonpractice disciplines (i.e., English, history, psychology)] were found in previous study, with nursing students among the highest scores (Walsh & Hardy, 1999). Another study found that average score of CTDI-CV and *analycity* in nursing were higher than those of medical students (Ling, Yaqing, Ying, Ping, & Li-sha, 2010). Due to the imbalance of sample size of nursing versus medical students, further studies are needed to explore the differences of CT dispositions among majors.

There are limitations of the current study, which await further explorations. First, this inventory was based on convenient sampling in one medical university of China, it is not clear that the sample's gender imbalance necessarily precluded a comparison of medical specialties. So a larger sampling covering more sites is needed in future studies. *Second*, direct comparison among three Chinese versions of critical thinking disposition inventory should be considered. Third, as most studies in critical thinking disposition, the current study was descriptive without analyzing causes of differentiated critical thinking dispositions across cultures, such as teaching and learning strategies. Further study may utilize active learning approaches such as problem-based learning (PBL) (Ozturk, Muslu, & Dicle, 2008; Yu, Zhang, Xu, Wu, & Wang, 2013), and intervention programs targeting motivation components such as self-awareness and mindfulness (Krupat et al., 2011) to confirm the effectiveness and validity of CDTI-CM. Last but not the least, the self-reported disposition may be subject to demand characteristics and social desirability, which is common to CT disposition scales. Future studies demand for development of more reliable test such as behavioral and cognitive tasks (e.g. cognitive reflection test, which we presumed to measure *analyticity*), and comparison between these different measures of critical thinking.

***Acknowledgements***

This study was supported by a grant from the Medical Education Branch of Chinese Medical Association and the Medical Education Professional Committee of Chinese Association of Higher Education in 2010 (2010-02-33) and Project of teaching reform research of Army Medical University (2014B06) and Project of Innovation and Entrepreneurship Training for College Students (201590031003). We are deeply grateful for the comments of Dr. Mark Battersby to the manuscript. And we also appreciate for the helpful discussions with Dr. Dongping Gao during preparation of the manuscript.

# References:

Aizikovitsh, E., Amit, M. (2010). Evaluating an infusion approach to the teaching of critical thinking skills through mathematics. *Procedia - Social and Behavioral Sciences, 2*(2), 3818-3822.

Aizikovitsh-Udi, E., Radakovic, N. (2012). Teaching Probability by Using Geogebra Dynamic Tool and Implemanting Critical Thinking Skills. *Procedia - Social and Behavioral Sciences, 46*(0), 4943-4947.

Albergaria-Almeida, P. (2011). Critical Thinking, Questioning and Creativity as Components of Intelligence. *Procedia - Social and Behavioral Sciences, 30*, 357-362.

Athari, Z. S., Sharif, S. M., Nasr, A. R., Nematbakhsh, M. (2013). Assessing critical thinking in medical sciences students in two sequential semesters: Does it improve? *J Educ Health Promot, 2*, 5.

Dwyer, C. P., Hogan, M. J., Stewart, I. (2014). An integrated critical thinking framework for the 21st century. *Thinking Skills and Creativity, 12*, 43-52.

Fábián, G. N. (2015). Non-critical thinking: What if not thinking? *Procedia - Social and Behavioral Sciences, 186*, 699-703.

Facione, N. C., Facione, P. A., Sanchez, C. A. (1994). Critical thinking disposition as a measure of competent clinical judgment: the development of the California Critical Thinking Disposition Inventory. *J Nurs Educ, 33*(8), 345-350.

Gupta, M., Upshur, R. (2012). Critical thinking in clinical medicine: what is it? *J Eval Clin Pract, 18*(5), 938-944.

Hajhosseiny, M. (2012). The Effect of Dialogic Teaching on Students' Critical Thinking Disposition. *Procedia - Social and Behavioral Sciences, 69*, 1358-1368.

Harasym, P. H., Tsai, T., Hemmati, P. (2008). Current Trends in Developing Medical Students' Critical Thinking Abilities. *The Kaohsiung Journal of Medical Sciences, 24*(7), 341-355.

Jenicek, M., Croskerry, P., Hitchcock, D. L. (2011). Evidence and its uses in health care and research: the role of critical thinking. *Med Sci Monit, 17*(1), A12-A17.

Jenkins, S. D. (2011). Cross-cultural perspectives on critical thinking. *J Nurs Educ, 50*(5), 268-274.

Kenimer, E. A. (2002). The identification and description of critical thinking behaviors in the practice of clinical laboratory science, Part 1: Design, implementation, evaluation, and results of a national survey. *J Allied Health, 31*(2), 56-63.

Kermansaravi, F., Navidian, A., Kaykhaei, A. (2013). Critical Thinking Dispositions Among Junior, Senior and Graduate Nursing Students in Iran. *Procedia - Social and Behavioral Sciences, 83*(0), 574-579.

Krupat, E., Sprague, J. M., Wolpaw, D., Haidet, P., Hatem, D., O'Brien, B. (2011). Thinking critically about critical thinking: ability, disposition or both? *Med Educ, 45*(6), 625-635.

Ku, K. Y. L., Ho, I. T. (2010). Dispositional factors predicting Chinese students' critical thinking performance. *Personality and Individual Differences, 48*(1), 54-58.

Li, L., Wegerif, R. (2014). What does it mean to teach thinking in China? Challenging and developing notions of 'Confucian education'. *Thinking Skills and Creativity, 11*(0), 22-32.

Li, X., Li, X., Lǜ, A. (2006). A comparative study on critical thinking ability of college nursing students in China, Japan and Samoa. *CHINESE NURSING RESEARCH, 20*(6B), 1521-1523.

Ling, B., Yaqing, Z., Ying, C., Ping, W., Li-sha, L. (2010). Analysis on Current Status and Related Factors of Critical Thinking Ability in Nursing and Med-ical Undergraduates. *Chinese Journal of School Doctor, 9B*(27), 1369-1372.

Luo, Q., Yang, X. (2001). Revision for CCTDI (Chinese Version). *Psychological Development and Education*(3), 47-51.

Nisbett, R. E. (2005). The influence of culture: holistic versus analytic perception.*, Trends in Cognitive Sciences 9(10) pp. 467 - 473*.

Ozturk, C., Muslu, G. K., Dicle, A. (2008). A comparison of problem-based and traditional education on nursing students' critical thinking dispositions. *Nurse Educ Today, 28*(5), 627-632.

Papp, K. K., Huang, G. C., Lauzon, C. L., Delva, D., Fischer, M., Konopasek, L., et al. (2014). Milestones of Critical Thinking: A Developmental Model for Medicine and Nursing. *Acad Med*.

Peng, K., Nisbett, R. E. (1999). Culture, Dialectics, and Reasoning About Contradiction. *American Psychologist, 54*(9), 741.

Peng, M., Wang, G., Chen, J. (2004). validity and reliability of the Chinese Critical Thinking Disposition Inventory. *Chinese Journal of Nursing, 39*(9), 644-647.

Salsali, M., Tajvidi, M., Ghiyasvandian, S. (2013). Critical thinking dispositions of nursing students in Asian and non-Asian countries: a literature review. *Glob J Health Sci, 5*(6), 172-178.

Shiau, S. J., Chen, C. H. (2008). Reflection and critical thinking of humanistic care in medical education.

*Kaohsiung J Med Sci, 24*(7), 367-372.

Shirkhani, S. 和. F. M. (2011). Enhancing Critical Thinking In Foreign Language Learners. *Procedia - Social and Behavioral Sciences, 29*, 111-115.

Spencer, C. (2008). Critical thinking in nursing: Teaching to diverse groups. *Teaching and Learning in Nursing, 3*(3), 87-89.

Stern, D. T., Wojtczak, A., Schwarz, M. R. (2002). Global minimum essential requirements in medical education. *Medical Teacher, 24*(2), 130-135.

Tang, T. (2004). Thought Discrepancies - On Language and Culture between Chinese and Westerners. *JOURNAL OF HAINAN NORMAL UNIVERSITY, 17*(72), 93-96.

Tiwari, A., Avery, A., Lai, P. (2003). Critical thinking disposition of Hong Kong Chinese and Australian nursing students. *J Adv Nurs, 44*(3), 298-307.

Walsh, C. M., Hardy, R. C. (1999). Dispositional differences in critical thinking related to gender and academic major. *J Nurs Educ, 38*(4), 149-155.

Wang, X., Sun, H., Zhang, A., Sun, W., Wang, P., Wang, Z. (2011). Potential role of metabolomics apporoaches in the area of traditional Chinese medicine: As pillars of the bridge between Chinese and Western medicine. *Journal of Pharmaceutical and Biomedical Analysis, 55*(5), 859-868.

Wojtczak, A., Schwarz, M. R. (2003). Global Minimum Essential Requirements: road to competence-oriented assessment of medical education programs. *Educación Médica*.

Xie, Z., Cao, W., Zhang, W., Zhu, H., Zhang, W. (2014). Critical thinking dispositions and related factors among medical undergraduates. *China Higher Medical Education* (10), 31-32.

Yeh, M. L. (2002). Assessing the reliability and validity of the Chinese version of the California Critical Thinking Disposition Inventory. *Int J Nurs Stud, 39*(2), 123-132.

Yeh, M. L., Chen, H. H. (2003). Comparison affective dispositions toward critical thinking across Chinese and American baccalaureate nursing students. *J Nurs Res, 11*(1), 39-46.

Yu, D., Zhang, Y., Xu, Y., Wu, J., Wang, C. (2013). Improvement in critical thinking dispositions of undergraduate nursing students through problem-based learning: a crossover-experimental study. *J Nurs Educ, 52*(10), 574-581.